\documentclass[12pt]{article}
\usepackage{cite}

\def\beq{\begin{equation}}
\def\eeq{\end{equation}}
\def\bea{\begin{eqnarray}}
\def\eea{\end{eqnarray}}

\begin{document}

\thispagestyle{empty}
\vspace*{.5cm}
\noindent
DESY 04-130 \hspace*{\fill} July 22, 2004\\
\vspace*{1.6cm}

\begin{center}
{\Large\bf A Comment on the Cosmological Constant Problem\\[0.3cm]
in Spontaneously Broken Supergravity}
\\[2.0cm]
{\large A. Hebecker}\\[.5cm]
{\it Deutsches Elektronen-Synchrotron, Notkestra\ss e 85, D-22603 Hamburg,
Germany}\\[.4cm]
{\small\tt (\,arthur.hebecker@desy.de\,)}
\\[1.1cm]

{\bf Abstract}\end{center}
\noindent
In spontaneously broken supergravity with non-flat potential the 
vanishing of the cosmological constant is usually associated with
a non-trivial balancing of two opposite-sign contributions. We make the 
simple observation that, in an appropriately defined expansion of the 
superfield action in inverse powers of $M_P$, this tuning corresponds to the 
absence of two specific operators. It is then tempting to speculate what 
kind of non-standard symmetry or structural principle might underlie the 
observed extreme smallness of the corresponding coefficients in the real 
world. Independently of such speculations, the suggested expansion appears 
to be a particularly simple and convenient starting point for the 
effective field theory analysis of spontaneously broken supergravity models.

\newpage
Supersymmetry or, more specifically, supergravity has so far not been 
able to offer a solution to the fundamental problem of the smallness of the 
cosmological constant. In fact, when compared with a simple non-SUSY model 
of gravity and a scalar field $\phi$, the situation might even 
look worse in the following sense. Expanding the non-SUSY lagrangian 
in inverse powers of the Planck mass $M_P$ (and in powers of $\phi$, as
far as the scalar field part is concerned), one finds that the cosmological 
constant problem is characterized by the anomalous smallness of the 
coefficient of det($g_{\mu\nu}$). However, even in the simplest version 
of spontaneously broken supergravity, involving just one chiral matter 
superfield, one apparently has to balance a positive and a negative 
contribution to the potential (cf.~Eq.~(\ref{potc}) below). At the minimum 
of the potential, these two contributions, each of which is derived in a 
non-trivial way from the fundamental action, have to cancel with very high 
precision. The only known cases where this cancellation is achieved for 
structural reasons, the so-called no-scale models, suffer from the 
presence of an exactly massless scalar field. 

In this short note we point out that it is possible to define an expansion 
of the supergravity action in inverse powers of $M_P$ in such a way that 
the requirement of a SUSY-breaking vacuum with a vanishing cosmological 
constant amounts simply to the absence of two specific operators. In short, 
using a K\"ahler-Weyl frame where the superpotential is constant and 
expanding the function exp$\,K(\Phi,\bar{\Phi})$ in powers of the chiral 
superfield $\Phi$, one has to forbid the terms $\Phi\bar{\Phi}$ and 
($\Phi^2\bar{\Phi}+$h.c.). As expected, there is no obvious symmetry 
principle enforcing this form, but a certain suggestive structure excluding 
the two dangerous operators will be identified below. On the one hand, these 
observations are extremely simple and their value might be merely 
pedagogical. On the other hand, one might hope that a deeper structural 
understanding of supergravity and its UV completion will emerge and justify 
the specific constraints mentioned above and to be discussed in more detail 
in the following. 

To be more specific, we start with the superfield action for a supergravity 
model~\cite{sugra} with one chiral superfield $\Phi$ as it emerges from the 
curved superspace approach (see e.g.~\cite{css,sga}). We follow the 
formulation 
of~\cite{bk}, which is very close to the more condensed treatment 
of~\cite{wb} (see also~\cite{ggrs}), and write the action as 
\beq
S=\int d^4x\,d^2\theta\,d^2\bar{\theta}\,E\left\{\Omega(\Phi,\bar{\Phi})+
\left(\frac{1}{R}W(\Phi)+\mbox{h.c.}\right)\right\}\,.\label{act}
\eeq
Here $E$ is the supervierbein determinant, $R$ is the superspace curvature, 
$W$ is the 
superpotential and $\Omega$, called the `superspace kinetic energy' 
in~\cite{wb}, is related to the K\"ahler potential by $K=-3\ln(-\Omega/3 
M^2)$ with the reduced Planck mass $M=M_P/\sqrt{8\pi}$. The field $\Phi$ 
is covariantly chiral, $\bar{\cal D}_{\dot{\alpha}}\Phi=0$, and the same is 
true for $R$. 

In flat space and setting all fermionic fields to zero, the above action 
takes the form
\beq
S=\int d^4x\,d^2\theta\,d^2\bar{\theta}\varphi\bar{\varphi}\Omega(\Phi, 
\bar{\Phi})+\left(\int d^2\theta\varphi^3W(\Phi)+\mbox{h.c.}\right)+\cdots
\,.\label{simp}
\eeq
where $\Phi$ is now a flat-space chiral superfield. This simple form is made 
possible by introducing the chiral compensator~\cite{sga,bk,wb,ggrs,kl,cfgp, 
ku} which, again in flat space and without fermions, is given (in 
`gravitational superfield gauge'~\cite{bk}) by $\varphi=1+\theta^2F_\varphi$. 
The ellipses in Eq.~(\ref{simp}) stand for extra kinetic terms for the scalar 
component of $\Phi$ which originate from integrating out the auxiliary vector 
field and have no simple flat-superspace representation. 

It is now easy to eliminate $F_\Phi$ and $F_\varphi$ and to obtain the scalar
potential. Before doing so, we make the additional assumption that the
superpotential is constant, $W(\Phi)=c_WM^3$. At the moment, this can be done 
without loss of generality since, given that $W$ is non-zero in the vacuum, 
it can always be made constant by performing an appropriate K\"ahler-Weyl 
transformation. However, later on we will have to impose constraints on 
$\Omega$ and this simple basis choice will be promoted to a true constraint. 

Thus, for $W(\Phi)=c_WM^3$, the scalar potential derived from Eq.~(\ref{simp})
reads
\beq
V_{\rm non-can.}(\Phi,\bar{\Phi})=\frac{9|c_W|^2M^6\Omega_{\Phi\bar{\Phi}}}
{\Omega_{\Phi\bar{\Phi}}\Omega-|\Omega_\Phi|^2}\,,\label{pot}
\eeq
where, as is commonly done, we use the same symbol for the superfield $\Phi$ 
and its scalar component. The index `non-can.' serves as a reminder that 
metric and scalar field are not canonically normalized. In particular, 
the Ricci-scalar enters the action with the coefficient $\Omega/6$. The 
indices of $\Omega$ symbolize partial derivatives.

Of course, one could have found an equivalent result by starting from the 
well-known formula (valid for canonical field normalization)
\beq
V(\Phi,\bar{\Phi})=\frac{1}{M^2}\exp(K)\left((K_{\Phi\bar{\Phi}})^{-1}
\left|W_\Phi+K_\Phi W\right|^2-3|W|^2 \right)\,,\label{potc}
\eeq
assuming constant $W$, and expressing $K$ through $\Omega$. The utility of 
such a formulation has been recognized and exploited for a detailed 
analysis of positivity properties of the potential in~\cite{bcf} (see 
also~\cite{aafl}). We note that $V=(3M^2/\Omega)^2V_{\rm non-can.}$. 

The crucial point now comes with the assumption that at $\Phi=0$ (which, 
as will be argued later, may be somehow distinguished at a fundamental 
level) the potential and its first derivatives vanish. In other words,
the point $\Phi=0$ is a SUSY-breaking vacuum with zero 
cosmological constant. We want to analyse this assumption in the spirit of
low-energy effective field theory, i.e., expanding the action of 
Eq.~(\ref{act}) (and thus the function $\Omega$) in powers of $\Phi/M$: 
\bea
\Omega(\Phi,\bar{\Phi})&=&-3M^2+(c_\Phi\Phi M+\mbox{h.c.})+c_{\Phi\bar{\Phi}}
\Phi\bar{\Phi}+(c_{\Phi^2}\Phi^2+\mbox{h.c.})\nonumber\\
\nonumber\\
&&+\left(c_{\Phi^2\bar{\Phi}}\frac{\Phi^2\bar{\Phi}}{M}+\mbox{h.c.}\right)+
\left(c_{\Phi^3}\frac{\Phi^3}{M}+\mbox{h.c.}\right)+\cdots\,.\label{om}
\eea
It can now be easily seen from Eq.~(\ref{pot}) that the above two 
assumptions simply mean that the coefficients $c_{\Phi\bar{\Phi}}$ and 
$c_{\Phi^2\bar{\Phi}}$ 
vanish. The constant term in Eq.~(\ref{om}) is used to define the Planck 
scale. The linear term is crucial for SUSY breaking and would, if the series 
were to end there, define a no-scale model~\cite{ns}. (This would, of 
course, also be true if the series were to continue with only holomorphic 
and antiholomorphic terms.) The term $\Phi\bar{\Phi}$ must be zero to have 
vanishing vacuum energy and, given that this is the case, it is now only 
the term $\sim\Phi^2\bar{\Phi}$ which could produce a linear term in the 
scalar potential. Thus, it has to be forbidden as well, giving the scalar 
potential
\beq
V(\Phi,\bar{\Phi})=-\frac{9|c_W|^2M^2}{|c_\Phi|^2}\left\{4c_{\Phi^2
\bar{\Phi}^2}\Phi\bar{\Phi}+(3c_{\Phi^3\bar{\Phi}}\Phi^2+\mbox{h.c.})+{\cal 
O}(|\Phi|^3/M)\right\}\,.
\eeq
This defines the scalar masses in terms of $c_W$, $c_\Phi$ and the 
coefficients $c_{\Phi^2\bar{\Phi}^2},\,\,c_{\Phi^3\bar{\Phi}}$, which we 
assume to have values making the extremum at $\Phi=0$ stable. The gravitino 
mass is given by $|c_W|M$. 

In summary, we have assumed that the action of Eq.~(\ref{act}), with $W$ 
being constant, gives rise to a flat, SUSY-breaking vacuum at $\Phi=0$. 
Expanding around this point in inverse powers $M$ (which is defined by the 
coefficient of the $\Phi$-independent term of $\Omega$), we have found that 
two of the coefficients have to vanish for consistency. Then we have 
identified gravitino and scalar masses in terms of the remaining 
coefficients. It should also be noted that, when deriving Eq.~(\ref{pot}) 
in this setting, $F_\varphi$ is found to be zero in the vacuum.

We conclude with some comments concerning possible structural reasons for the 
absence of the two offending terms and on the generalization to more than 
one chiral matter superfield. It is easy to see that a `superspace kinetic 
energy' of the 
form 
\beq
\Omega(\Phi,\bar{\Phi})=f(\Phi)+f(\bar{\Phi})+F(\Phi^2,\bar{\Phi}^2)
\label{oa} 
\eeq
automatically has vanishing $c_{\Phi\bar{\Phi}}$ and $c_{\Phi^2\bar{\Phi}}$. 
(Our notation implies that 
$f$ and $F$ are defined as power series of their respective arguments.) One 
way to look at this is to view $F$ as a very specific correction to a 
no-scale model, ensuring that the flatness is lifted while the zero vacuum
energy is preserved. Another possible attitude is to consider $f$ as a 
correction to a model defined by the completely generic `kinetic function' 
$F(\Psi,\bar{\Psi})$. This correction would have to be introduced in a very 
peculiar way, namely by adding the no-scale type function $f(\Phi)+f(\bar{ 
\Phi})$ after the reparameterization $\Psi\to\Phi^2$. At the moment, neither 
of these possibilities may appear particularly convincing or easy to 
justify at the quantum level. However, we recall that a similar criticism 
may be applied, for example, to the ansatz of writing both superpotential 
and `kinetic function' as the sum of hidden and visible sector 
contributions~\cite{rs}. Yet, this turns out to be natural if the two 
sectors are separated in higher dimensions. In this spirit, one may 
hope that the ansatz of Eq.~(\ref{oa}) or some similar ansatz justifying 
$c_{\Phi\bar{\Phi}}=c_{\Phi^2\bar{\Phi}}=0$ will eventually find a deeper 
understanding. 

An obvious way to generalize the above discussion to 
more superfields is by adding a set of chiral fields $Q=\{Q^i\}$ which 
transform under some global, continuous, unitary symmetry. It is easy to 
see that, making all coefficients $c$ in Eq.~(\ref{om}) and $c_W$ arbitrary 
functions of $Q$ (respecting, of course, the symmetry) will not affect the 
existence of a SUSY-breaking stationary point at $\Phi=0$ and $Q=0$ with
vanishing cosmological constant. The reason is the enhanced symmetry at 
this point, which makes it impossible to have linear terms driving either 
$Q$ or $F_Q$ to non-zero values. Of course, to make this point a 
minimum rather than just an extremum imposes further conditions on the 
coefficients in the lagrangian which, however, do not involve any 
fine-tuning. 

A more interesting generalization is to include further uncharged 
fields, so that all fields in the set $\{\Phi^i\}$ can contribute linear 
terms to $\Omega$ and mix with each other. Following the one-field case 
discussed above, we expand $\Omega$ around $\Phi^i=0$ (for all $i$) and 
require the presence of a linear contribution. As has been shown 
in~\cite{bcf} (cf.~also~\cite{lps}), the scalar potential is proportional to 
det$[\Omega_{i \bar{j}}(\Phi,\bar{\Phi})]$. 
(Here the indices indicate derivatives with 
respect to $\Phi^i$ and $\bar{\Phi}^j$.) The vanishing of the cosmological 
constant at $\Phi^i=0$ corresponds to det$[\Omega_{i\bar{j}}(0,0)]=0$, which
is the generalization of the above condition $c_{\Phi\bar{\Phi}}=0$ to the 
multi-field 
case. The vanishing determinant of $\Omega_{i\bar{j}}(0,0)$ implies the 
existence of a vector $v^{\bar{i}}$ annihilated by this hermitian matrix, 
and we can choose a basis where only the first component of this vector is 
different from zero. In this basis, one has $\Omega_{1\bar{i}}(0)=\Omega_{i 
\bar{1}} (0)=0$. This implies that the first derivative of 
det$[\Omega_{i\bar{j}}]$
at the point $\Phi^i=0$ can only be non-zero if the first derivative of 
the element $\Omega_{1\bar{1}}$ is non-zero at this point. Thus, given the 
above basis choice, the stationarity condition (the previous condition 
$c_{\Phi^2\bar{\Phi}}=0$) generalizes to $\Omega_{1\bar{1}i}(0,0)=0$ in the 
multi-field 
case. This condition can also be given in the basis independent form 
$v^i\Omega_{i\bar{i}j}(0,0)v^{\bar{i}}=0$. Thus, it appears that a nonlinear 
relation between different coefficients in the expansion of $\Omega(\Phi, 
\bar{\Phi})$ has arisen. The simple form of the constraints found in the 
one-field case does not emerge unless one distinguishes a specific field
which, as an a-priori feature, does not participate in any non-holomorphic 
second-order terms in $\Omega$. 

\noindent
{\bf Acknowledgements}: I would like to thank W.~Buchm\"uller, S.~Ferrara, 
and R.~Sundrum for useful comments and suggestions. I am also grateful to 
B.~Wonsak for many discussions during the preparation of his Diploma 
thesis~\cite{won}.

\end{document}